\documentclass[a4paper,fleqn,usenatbib]{mnras}

\usepackage{newtxtext,newtxmath}

\usepackage[T1]{fontenc}
\usepackage{ae,aecompl}

\usepackage{graphicx}	
\usepackage{amsmath}	
\usepackage{amssymb}	

\def\teff{$T\rm_{eff}$}
\def\kms{$\mathrm{km\, s^{-1}}$}

\def\vt{V$_{\rm t}$}
\def\logg{$\log\,g$}

\title[Copper abundance from \ion{Cu}{i} and \ion{Cu}{ii} lines in metal-poor star]{
Copper abundance from \ion{Cu}{i} and \ion{Cu}{ii} lines in metal-poor star spectra: 
NLTE vs LTE.\
\thanks{Based on observations taken at ESO, programmes 
65.L-0507,
72.B-0179, 
72.B-0585, 
74.B-0639, 
76.B-0133, 
266.D-5655 and HST programme
GO-7348,
GO-8197, 
GO-9804, 
GO-14161,
GO-14672
}
}

\author[S.~Korotin et al.]{
S.A.~Korotin,$^{1}$\thanks{E-mail: serkor@skyline.od.ua},
S.M. Andrievsky$^{2,3}$
and A.V. Zhukova$^{1}$
\\
$^{1}$Crimean Astrophysical Observatory, Nauchny 298409, Republic of Crimea\\
$^{2}$Department of Astronomy and Astronomical Observatory, Odessa National University, Isaac Newton Institute of Chile, Odessa Branch,\\ 
Shevchenko Park, 65014, Odessa, Ukraine\\
$^{3}$GEPI, Observatoire de Paris, PSL Research University, CNRS, Place Jules Janssen, 92195 Meudon, France\\
}

\date{Accepted XXX. Received YYY; in original form ZZZ}

\pubyear{2018}

\begin{document}
\label{firstpage}
\pagerange{\pageref{firstpage}--\pageref{lastpage}}
\maketitle

\begin{abstract}

We checked consistency between the copper abundance derived in six
metal-poor stars using UV \ion{Cu}{ii} lines (which are assumed to form in LTE) 
and UV \ion{Cu}{i} lines (treated in NLTE). Our program stars cover
the atmosphere parameters which are typical for intermediate temperature dwarfs
(effective temperature is in the range from approximately 5800 to 6100 \,K,
surface garvity is from 3.6 to 4.5, metallicity is from about --1 to --2.6 dex). 
We obtained a good agreement between abundance from these two sets of the lines, 
and this testifies about reliability of our NLTE copper atomic model. We confirmed 
that no underabundace of this element is seen at low metallicities (the mean [Cu/Fe]
value is about --0.2 dex, while as it follows from the previous LTE studies copper behaves
as a secondary element and [Cu/Fe] ratio in the range of [Fe/H from --2 to --3 dex should 
be about --1 dex). According to our NLTE data the copper behaves as a primary element 
at low metallicity regime. We also conclude that our new NLTE copper abundance 
in metal-poor stars requires significant reconsideration of this element yields in the 
explosive nucleosynthesis.

\end{abstract}




\section{Introduction}

Copper is the rather problematic element from the point of view of the Galaxy 
chemodynamical model. Several studies on the copper abundance determination 
were based on the use of the green subordinate \ion{Cu}{i} lines  5105.5~\AA, 
5218.2~\AA, 5782.1~\AA. According to the papers published by \citet{Cohen80}, 
\citet{Sneden91}, \citet{Mishenina02}, \citet{Simmerer}, 
\citet{Bihain}, \citet{Bonifacio10} the copper-to-iron ratio decreases 
with iron abundance decrease. Specifically, \citet{Cohen80} derived abundances 
in sreveral globular clusters. Copper abundance (based on 5782.1~\AA\ line) showed 
progressively decreasing content as the cluster metallicity was decreasing. 
\citet{Sneden91} derived copper abundance in a sample of the 
field and globular cluster metal-poor stars ([M/H] from --2.5 to --1) using 
two lines: 5105.5~\AA\ and 5782.1~\AA. They conclude that copper abundance 
behaviour in metal-poor stars may be explained if one suppose that there is 
a weak $s$-process component that mainly contributes to the copper production, 
while exposive nucleosynthesis has a secondary influence on the copper cosmic 
production. \citet{Mishenina02} studied 90 metal-poor stars 
([M/H] from --3 to -- 0.5) and found that abundance of the copper derived from 
the green 5105.5~\AA, 5218.2~\AA\ and 5782.1~\AA\ lines shows a clear 
decrease of the [Cu/Fe] ratio towards the lower iron content. 
Generally, the results of \citet{Mishenina02} confirmed conclusion of 
\citet{Sneden91} about the rather sharp decrease of the relative 
copper abundance at metallicities of about --1.5.  \citet{Simmerer} 
analyzed 117 giants in ten globular clusters using two optical lines 
5105.5~\AA\ and 5782.1~\AA. They found the same trend of [Cu/Fe] decrease 
with [Fe/H] decrease as \citet{Sneden91} and \citet{Mishenina02} did.
Generally, with a metallicity decreasing the relative copper abundance is 
gradually decreasing and achieving [Cu/Fe] $\approx -0.8$ at [Fe/H] = --2. 
In the range of the lower metal content the [Cu/Fe] ratio achieves of some 
plateau value from --0.8 to --1.0 dex.

To reproduce observed by  \citet{Sneden91} copper underabundance in 
the domain of metal-poor stars \citet{TWW95} decreased the iron 
yields from SNe II. \citet{Romano10} stress an importance of SNe II as 
the copper nuclei source for the short phase  in the beginning
of the Galaxy evolution. \citet{Matteucci} discussed constraints on the 
nucleosynthesis of copper basing on the observational data of
\citet{Sneden91}. They make conclusion that copper is produced via several 
processes: $s$-process in massive stars (weak component), $s$-process in low 
mass stars (main component), as well as in explosive nucleosynthesis in SNe II 
and SNe Ia. In order to achieve an agreement between the model prediction and 
observed abundances those authors assumed that SNe Ia should start polluting 
the interstellar media already at [Fe/H] about --2. \citet{Mishenina02} argued 
that at low metallicity the great majority of the copper nuclei were 
produced  by secondary phenomena in massive stars and by SNe Ia on a long time 
scale. Thus, as one can see the SNe II are considered as an important source of this 
element production only at the very early stages of the Galaxy evolution. 
Summarizing, the observational results and results of determination of the copper abundances 
in metal-poor stars faced a problem in their interpretation and finding the proper 
site of this element production.

In past years it was a belief that optical copper lines are free of the NLTE 
effects (see, e.g. \citealt{Mishenina02}). The first indication that something 
could be wrong with this statement came from the paper of \citet{Shi14}, 
who noted that the important NLTE mechanism affecting the \ion{Cu}{i} 
spectrum is the UV overionization. Later \citet{Yan15} investigated copper 
abundance in 64 late-type intermediate metal-poor stars from Galactic disc 
and halo. The authors succeeded to show that NLTE effects are important 
for this atom (correction is of about 0.17 dex for metallicity --1.5). 
The NLTE calculations enhanced the obtained copper abundance in the range of 
metal-poor stars giving a more flatter distribution of [Cu/H] vs. [Fe/H] than 
it was declared in previous works based on LTE assumption. This conclusion was 
confirmed by \citet{Yan16} who stated that NLTE  effects are strong for the 
copper in metal-poor stars.

Up to now the most comprehensive  study of the copper Galactic evolution in the 
light of NLTE computation  was performed by \citet{Andr18}. In that 
paper it was shown that for the sample of intermediate and extreme metal-poor 
stars ([Fe/H] from  --4.2 to --1.4) the mean [Cu/Fe] value is about --0.22 dex. 
This is very different from results of LTE analyses which show significant 
underabundance at the early stages of the Galaxy evolution.

Very recently \citet{RB18} performed a new work on copper abundance in the 
late-type stars using UV \ion{Cu}{ii} lines which are supposedly free 
of the NLTE effect influence. HST and ground-based spectra were used for this 
aim. Authors showed that the mean copper abundance in six warm metal-poor 
([Fe/H] from --2.50 to --0.95) dwarfs derived from \ion{Cu}{i} and \ion{Cu}{ii} 
lines differs by 0.36 dex (\ion{Cu}{ii} lines give higher abundance comparing 
to \ion{Cu}{i} lines).  Generally, this difference agrees with derived NLTE 
corrections in metal-poor stars reported by \citet{Yan15} and \citet{Andr18}.

In this paper we examine NLTE copper abundance in a sample of the same
six metal poor stars derived from different \ion{Cu}{i} lines (including far UV region) 
and \ion{Cu}{ii} lines with the aim to get an independent estimate of the 
reliability of our NLTE corrections reported in \citet{Andr18}.  

\section{Spectroscopic material}

We employed UV stellar spectra that were secured with the help of Space 
Telescope Imaging Spectrograph (STIS; \citealt{Kimble}; \citealt{Woodgate}) 
on the board of Hubble Space Telescope (HST). Observations were made using the 
E230H Echelle grating, the 0.09" x 0.2" slit, and the NUV Multianode 
Microchannel Array detector. Resolved power was R=114 000.  
Investigated \ion{Cu}{i} and \ion{Cu}{ii} lines are situated in the range 
2024 -- 2248 \AA.

We retrieved archive spectra from Mikulski Archive for Space Telescopes (MAST).
All spectra are centered at 2013~\AA. The following lines fall in the spectral 
range: \ion{Cu}{i} 2024~\AA, \ion{Cu}{ii} 2037~\AA, 2055~\AA, 2104~\AA, 2112~\AA, 
2126~\AA\ (sometimes 2148~\AA\ line is available). In addition, for HD84937, 
HD94028, HD140283 there are the spectra in MAST that are centered on 2263~\AA. 
The following lines are available in those spectra: \ion{Cu}{i} 2165~\AA, 2199~\AA, 
2214~\AA, 2225~\AA, 2227~\AA, 2230~\AA\ and \ion{Cu}{ii}  2189~\AA, 2247~\AA. 
For HD140283 we also used spectra centered at 2113~\AA\ and 2163~\AA. It should be 
noted that S/N ratio is small (from 5 to 18). Since the number of available 
spectra (Nsp) varies from 3 to 22 (see Table 1) we co-added individual spectra and 
reached S/N ratio of about 25--50. This is enough to make reliable comparison between 
observed and synthesized profiles.

Optical \ion{Cu}{i} lines were investigated using the spectra from European Southern 
Observatory (ESO) Science Archive Facility. They were secured with the help of
the Ultraviolet and Visual Echelle Spectrograph (UVES; \citealt{Dekker}) on 
the Very Large Telescope. This enabled us to include in our consideration two 
resonant UV lines 3247~\AA\ and 3274~\AA. In spectra of two intermediate 
metal-poor stars (HD76932 and HD94028) the optical subordinate lines are 
seen. 

Spectroscopic data are listed in Table \ref{spectr}.

\begin{table}
\caption{UV and Optical Spectra.}
\label{spectr}
\begin{tabular}{lclcccc}
\hline
Star     & Instr. & Program ID & R      & $\lambda$(\AA) & Nsp & S/N\\
\hline
\multicolumn{2}{c}{UV Spectra}\\
\hline
HD 19445 &    STIS    & GO-14672   & 114000 & 1880--2150& 22& 50 \\
HD 76932 &    STIS    & GO-9804    & 114000 & 1880--2150&  9& 45 \\
HD 84937 &    STIS    & GO-14161   & 114000 & 1880--2150& 18& 30 \\
         &    STIS    & GO-14161   & 114000 & 2128--2404& 10& 35 \\
HD 94028 &    STIS    & GO-8197    & 114000 & 1880--2150& 13& 40 \\
         &    STIS    & GO-14161   & 114000 & 2128--2404&  9& 35 \\
HD140283 &    STIS    & GO-7348    & 114000 & 1930--2205&  4& 18 \\
         &    STIS    & GO-7348    & 114000 & 1980--2250&  4& 22 \\
         &    STIS    & GO-7348    & 114000 & 2128--2404&  3& 35 \\
         &    STIS    & GO-14161   & 114000 & 2128--2404&  3& 35 \\
HD160617 &    STIS    & GO-8197    & 114000 & 1880--2150& 15& 28 \\
\hline
\multicolumn{2}{c}{Optical Spectra}\\
\hline
HD 19445 &    UVES    & 074.B-0639 &  40970 & 3040--3870&  1& 130\\
HD 76932 &    UVES    & 266.D-5655 &  65030 & 3040--3870&  1& 120\\
         &    UVES    & 072.B-0179 & 107200 & 4620--6650&  1& 230\\
HD 84937 &    UVES    & 266.D-5655 &  65030 & 3040--3870&  1&  85\\
HD 94028 &    UVES    & 076.B-0133 &  36840 & 3040--3870&  1& 150\\
         &    UVES    & 072.B-0585 &  45254 & 4778--6807&  1& 150\\
HD140283 &    UVES    & 266.D-5655 &  65030 & 3040--3870&  1& 100\\
HD160617 &    UVES    & 065.L-0507 &  49620 & 3040--3870&  1& 250\\
\hline
\end{tabular}
\end{table}

\section{Copper Abundance Analysis}
Atmosphere parameters of the program stars were taken from 
\citet{RB18}. Those parameters are listed in Table \ref{param}. 
The atmosphere models were calculated with the help of ATLAS9 code and ODF 
from \citet{CK03}.

\begin{table}
\caption{Parameters of studied stars.}
\label{param}
\begin{tabular}{lccccc}
\hline
Star     &V    &\teff    &\logg      &\vt         & [Fe/H]\\
         &(mag)&(K)      &(cgs)      &(\kms)     &\\
\hline
HD 19445 &8.06 &6070$\pm$76&4.44$\pm$0.14&1.60$\pm$0.10& --2.12$\pm$0.05\\
HD 76932 &5.86 &5945$\pm$93&4.17$\pm$0.11&1.10$\pm$0.10& --0.95$\pm$0.06\\
HD 84937 &8.32 &6427$\pm$93&4.14$\pm$0.14&1.45$\pm$0.10& --2.16$\pm$0.05\\
HD 94028 &8.22 &6097$\pm$74&4.34$\pm$0.14&1.30$\pm$0.10& --1.52$\pm$0.05\\
HD140283 &7.22 &5766$\pm$64&3.64$\pm$0.13&1.30$\pm$0.10& --2.59$\pm$0.05\\
HD160617 &8.74 &6050$\pm$67&3.91$\pm$0.13&1.50$\pm$0.10& --1.89$\pm$0.04\\
\hline
\end{tabular}  
\end{table}

We enlarged the list of used \ion{Cu}{ii} lines by two times compared to that 
used by \citet{RB18}. In addition, in the UV region we considered from one to seven \ion{Cu}{i}
lines, the synthetic profiles of which we calculated taking into
account the deviations from LTE. For the resonant lines at 3247~\AA\ and 
3274~\AA\, as well as for the subordinate lines in the optical region we 
also used NLTE approximation. It should be noted that copper atoms are mainly remaining-
in the ionization state at the considered conditions. Therefore, the 
\ion{Cu}{ii} lines are supposedly free of the NLTE influence. At this time we 
cannot prove this supposition because of the lack of the atomic data for the 
\ion{Cu}{ii} ion.

In order to calculate the NLTE deviations in the level populations for 
\ion{Cu}{i} model we used MULTI code (\citealt{Car86}) modified by \citet{Kor99}. 
The \ion{Cu}{i} model is described in details in \citet{Andr18}. This model 
includes radiative and collisional transitions between 59 \ion{Cu}{i} atomic 
levels and the ground station of \ion{Cu}{ii} ion. 
Inelastic collisions with hydrogen atoms were described with the Drawin's 
formula \citet{Drawin68,Drawin69} adapted for astrophysical use by \citet{SH84}
without correction factor. The process of our NLTE copper atomic model 
testing with solar and stellar spectra is thoroughly described in \citet{Andr18}. 

Proper comparison of observed and computed profiles in many cases requires a 
multi-element synthesis to take into account possible blending lines of other 
species. For this process, we fold the NLTE (MULTI) calculations, specifically 
the departure coefficients, into the LTE synthetic spectrum code SYNTHV 
(\citealt{Tsym96}) that enables us to calculate the NLTE source function for 
copper lines. These calculations included all spectral lines from the VALD 
database (\citealt{Ryab15}) in a region of interest. The LTE approach 
was applied for lines other than the \ion{Cu}{i} lines. Abundances of corresponding 
elements were adopted in accordance with the [Fe/H] value for each star.

To fit the copper line profiles in the optical region we have taken into 
account the hyper-fine structure. The detailed list of the wavelengths and 
oscillator strengths is given in \citet{Andr18}. Table \ref{line} 
contains the mean wavelengths and averaged oscillator strengths.

We considered UV lines without hyper-fine structure. The oscillator strengths 
for \ion{Cu}{ii} lines were taken from \citet{Dong}, and for most neutral 
copper lines in UV -- from \citet{Kurucz11}. 
For line 2024 \AA\ we used data from  \citealt{Lindg80} and for the 
line 2165 \AA\ the data from \citealt{Morton91}. 
Oscillator strengths for the resonant \ion{Cu}{i} lines were determined with a 
high precision. Accordingly to the NIST database the error does not exceed 1\% 
(0.004 dex). For the subordinate lines the error value is not very high too 
(about 12 \% that corresponds to 0.05 dex). The error in the oscillator 
strengths for the UV lines is really increasing, but it does not exceed 18 \% 
(0.09 dex). The damping parameters were 
taken from the VALDatabase. All used data are listed in Table \ref{line}.
In the last column of this Table we list the van der Waals constant.

\begin{table}
\centering
\caption{UV and Optical Copper Lines.}
\label{line}
\begin{tabular}{llrc}
\hline
$\lambda$(\AA)& Elow(eV)& log gf& $\Gamma_{vw}$\\
\hline
\multicolumn{4}{c}{\ion{Cu}{i}}\\
\hline
2024.325& 0.0   & -1.75&  -7.47\\
2024.338& 0.0   & -1.46&  -7.47\\
2165.096& 1.3889& -0.84&  -7.81\\
2199.586& 1.3889&  0.45&  -7.68\\
2199.754& 1.6422&  0.34&  -7.46\\
2214.583& 1.3889&  0.11&  -7.31\\
2225.705& 0.0   & -1.20&  -7.81\\
2227.776& 1.6422&  0.46&  -7.63\\
2230.086& 1.3889&  0.64&  -7.58\\
\hline                       
3247.54& 0.0    &-0.05& -7.89\\
3273.95& 0.0    &-0.35& -7.89\\
5105.54& 1.3890 &-1.51& -7.72\\
5153.23& 3.7859 &-0.01& -7.28\\
5218.20& 3.8167 & 0.27& -7.28\\
5782.13& 1.6422 &-1.83& -7.82\\
\hline                       
\multicolumn{4}{c}{\ion{Cu}{ii}}\\
\hline
2037.127 & 2.8327& -0.28&  -7.91\\
2054.979 & 2.8327& -0.30&  -7.91\\
2104.796 & 2.9754& -0.60&  -6.16\\
2112.100 & 3.2564& -0.14&  -6.60\\
2126.044 & 2.8327& -0.32&  -6.62\\
2148.984 & 2.7188& -0.49&  -6.62\\
2189.630 & 3.2564& -0.39&  -6.61\\
2247.003 & 2.7188&  0.10&  -6.62\\
\hline                       
\end{tabular}
\end{table}

\section{Results and Discussion}
\begin{figure*}
\includegraphics[width=18cm,clip=true]{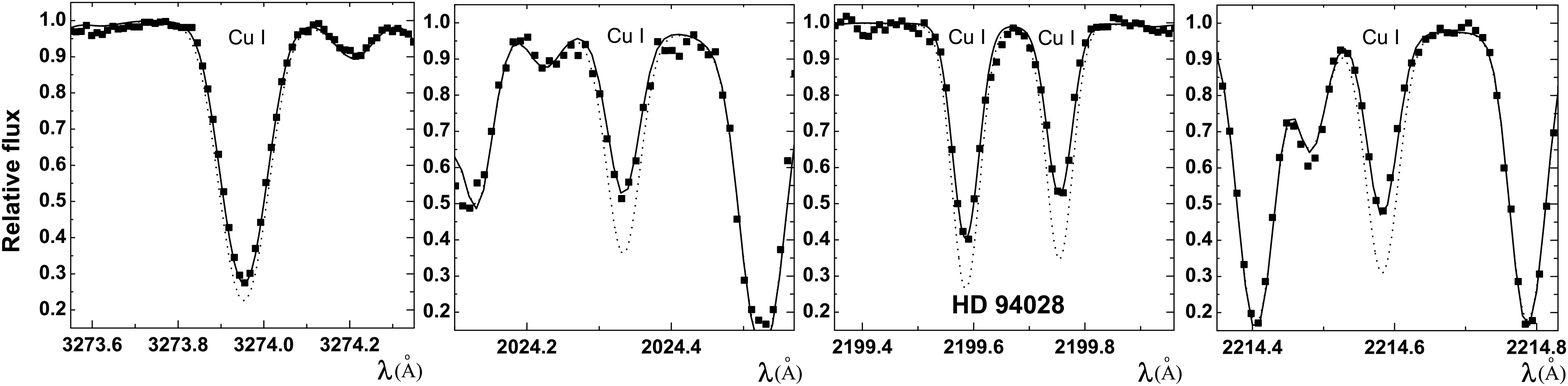}
\includegraphics[width=18cm,clip=true]{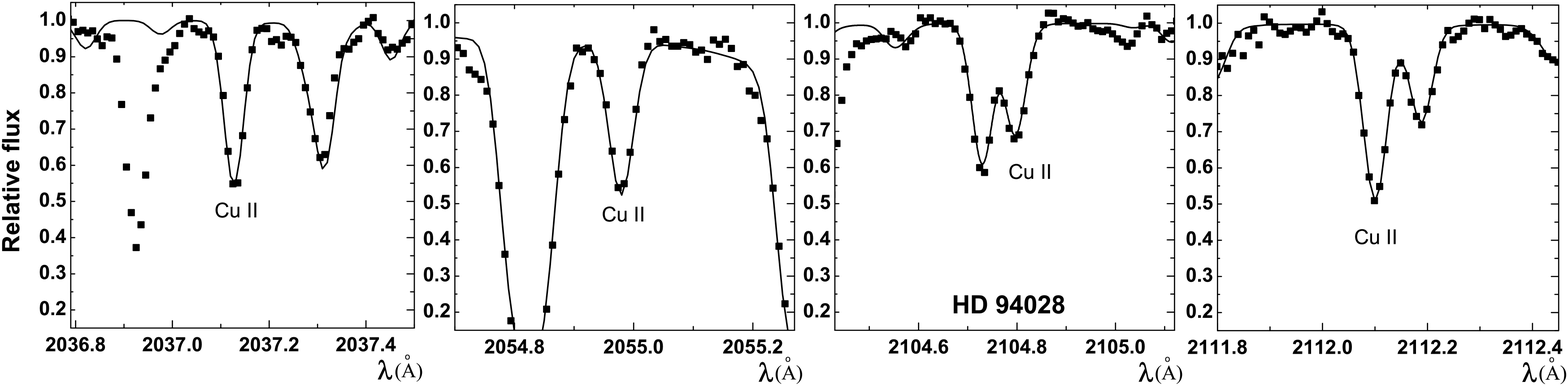}
\caption{The \ion{Cu}{i} lines in the spectrum of the HD\,94028 
(upper panel) and \ion{Cu}{ii} (lower panel), filled squares,
compared to our NLTE synthetic spectrum (solid) line and LTE synthetic spectrum 
(dotted line). }
\label{HD94028}
\end{figure*}

The copper abundance in our six program stars was derived by the \ion{Cu}{i} 
and \ion{Cu}{ii} line profile fitting. Those lines are situated in UV and 
optical regions. Abundance derived from \ion{Cu}{i} 3247 and 3274~\AA\ lines 
and UV lines of \ion{Cu}{i} agree well with those obtained from \ion{Cu}{ii} 
lines in UV region. This can be seen from the spectra fitting for 
HD 94028. The upper panel of the Fig. \ref{HD94028} shows the profiles of 
five \ion{Cu}{i} lines. The lower panel shows four \ion{Cu}{ii} line profiles. 
The LTE profiles of the \ion{Cu}{i} lines are indicated by the dotted lines 
(all of them were synthesized with the same abundance).

In Table \ref{res} we give the copper abundance that was derived from the 
profile fitting of individual lines in two ionization stages. The individual 
error in the abundance determination from the profile fitting varies from 
0.05 to 0.15 dex. The error in the mean copper abundance for each ionization 
stage is given in view of the accuracy of the profile fitting.
Therefore, despite of the close abundance values derived from the 
individual profiles, we estimate the mean abundance error as 0.05--0.12 dex.

\begin{table*}
\caption{Derived Abundances and Uncertainties for Individual Cu Lines}
\label{res}
\begin{tabular}{lccccccccccccccccccccccccccccccc}
\hline
\multicolumn{1}{c}{Star} & \multicolumn{11}{c}{$\log \epsilon$(Cu/H) + 12.00 from \ion{Cu}{i} lines (NLTE calculation)}  \\
\hline                                   
        &3247\AA&3274\AA&5105\AA&5153\AA&5578\AA&2024\AA&2165\AA&2199\AA&2214\AA&2225\AA&2227\AA&2230\AA&mean \\
\hline
HD 19445&1.81&1.81&    &    &    &1.73&    &    &    &    &    &    &1.77$\pm$0.08\\
HD 76932&3.29&3.29&3.33&3.37&3.38&3.35&    &    &    &    &    &    &3.34$\pm$0.05\\
HD 84937&1.83&1.83&    &    &    &1.83&1.77&1.90&1.83&1.82&1.83&1.79&1.83$\pm$0.12\\
HD 94028&2.52&2.52&2.62&    &    &2.62&2.70&2.60&2.62&2.70&2.62&2.58&2.62$\pm$0.10\\
HD140283&1.44&1.44&    &    &    &1.44&1.44&1.44&1.48&1.52&1.43&1.44&1.45$\pm$0.10\\
HD160617&2.09&2.09&    &    &    &2.08&    &    &    &    &    &    &2.09$\pm$0.08\\
\hline
\multicolumn{1}{c}{Star} & \multicolumn{11}{c}{$\log \epsilon$(Cu/H) +12.00 from \ion{Cu}{ii} lines (LTE calculation)}\\
\hline
        &2037\AA&2055\AA&2104\AA&2112\AA&2126\AA&2149\AA&2189\AA&2247\AA&mean&&&Cu~I--Cu~II&[Cu/Fe]\\
\hline                                                                                 
HD 19445&1.70&1.73&1.76&1.72&1.73&1.82&    &    &1.74$\pm$0.08&&&0.03&--0.37 \\
HD 76932&3.32&3.29&3.35&3.35&3.39&    &    &    &3.34$\pm$0.08&&&0.00& +0.04 \\
HD 84937&1.58&1.65&1.60&1.63&1.62&1.67&    &1.60&1.62$\pm$0.14&&&0.20&--0.37 \\
HD 94028&2.38&2.38&2.52&2.65&2.52&2.64&2.52&2.62&2.53$\pm$0.10&&&0.09&--0.15 \\
HD140283&1.42&1.44&1.44&1.43&1.39&1.44&1.44&1.35&1.42$\pm$0.15&&&0.04&--0.22 \\
HD160617&1.97&1.98&2.10&2.12&2.08&    &    &    &2.05$\pm$0.12&&&0.03&--0.29 \\
\hline
\end{tabular}
\end{table*}

We can state that our atomic model is correct since we derive the same 
abundance both from \ion{Cu}{i} (NLTE) and \ion{Cu}{ii} (LTE) lines. The 
only problem is the star HD 84937, but the difference between abundances is not
too big (about 0.2 dex). This can be explained by the uncertainties in its
atmosphere parameters. In fact, according to the different studies the 
effective temperature for this star differs from 6211 to 6541 K, surface 
gravity -- from 4.0 to 4.5 dex, \vt\ from 1.3 to 1.7 km/s. The corresponding
results were published in \citet{Prugniel11}, \citet{Battistini15}, 
\citet{Boeche16}, \citet{Mishenina17}, \citet{Peterson17}, \citet{Mashonkina17}. 
For the sake of a consistent comparison we used atmosphere parameters published 
in \citet{RB18}.

\begin{figure*}
\includegraphics[width=18cm,clip=true]{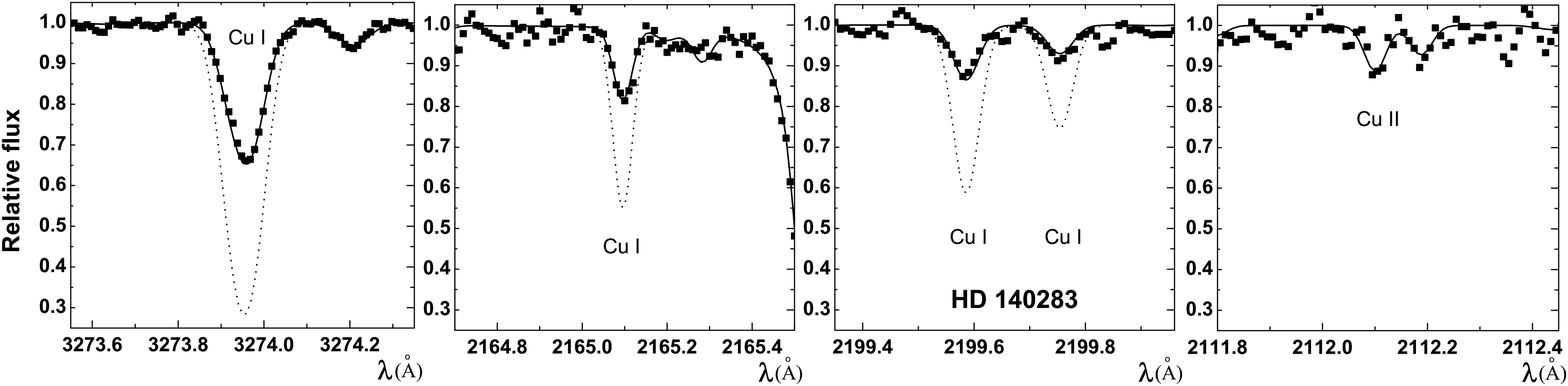}
\includegraphics[width=18cm,clip=true]{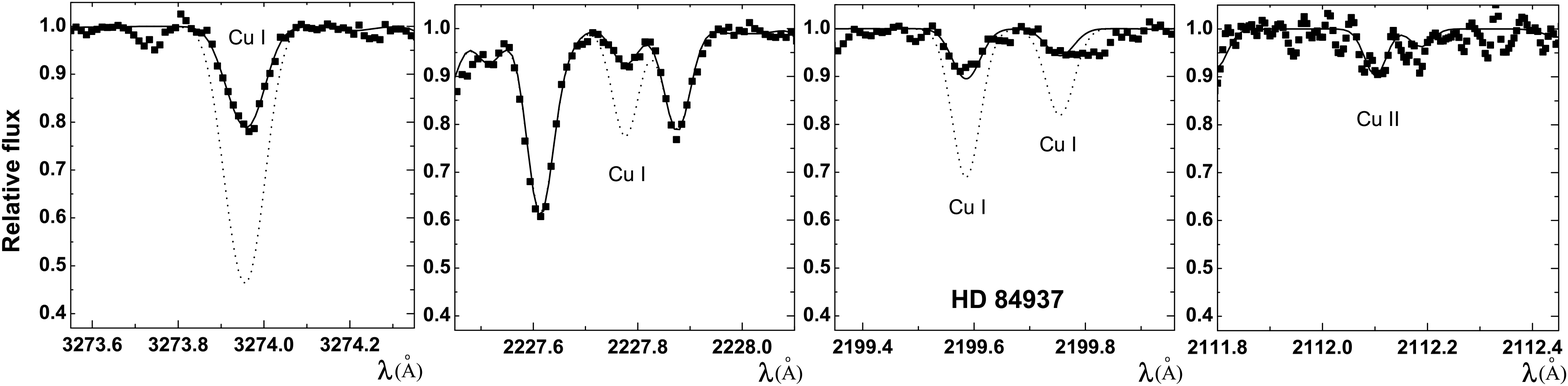}
\caption{The \ion{Cu}{i} 3273\,\AA, 2165\,\AA, 2199\,\AA, 
and \ion{Cu}{ii} 2112\,\AA\ lines in the spectrum of the HD\,140283, 
(upper panel), HD\,84937 (lower panel), filled squares,
compared to our NLTE synthetic spectrum (solid) line and LTE synthetic spectrum 
(dotted line). }
\label{HD140283_84937}
\end{figure*}

We should note that all the spectra in UV region have a quite low S/N ratio. 
Even a co-adding of all available spectra for a certain star does not allow to 
achieve the S/N ratio more than 35--50. In fact this hampers of getting the 
reliable abundance of the copper from \ion{Cu}{ii} lines. On contrary, the UV 
\ion{Cu}{i} lines in the range from $\lambda$ 2150 to 2230 \AA~ are stronger 
than \ion{Cu}{ii} and less affected by the low S/N ratio, and they give more 
reliable copper abundance. This is clearly seen in Fig. \ref{HD140283_84937}, 
where we show a comparison between observed and theoretical \ion{Cu}{i} and 
\ion{Cu}{ii} lines for two stars: HD 84937, HD 140283.

\begin{figure}
\includegraphics[width=0.9\columnwidth,clip=true]{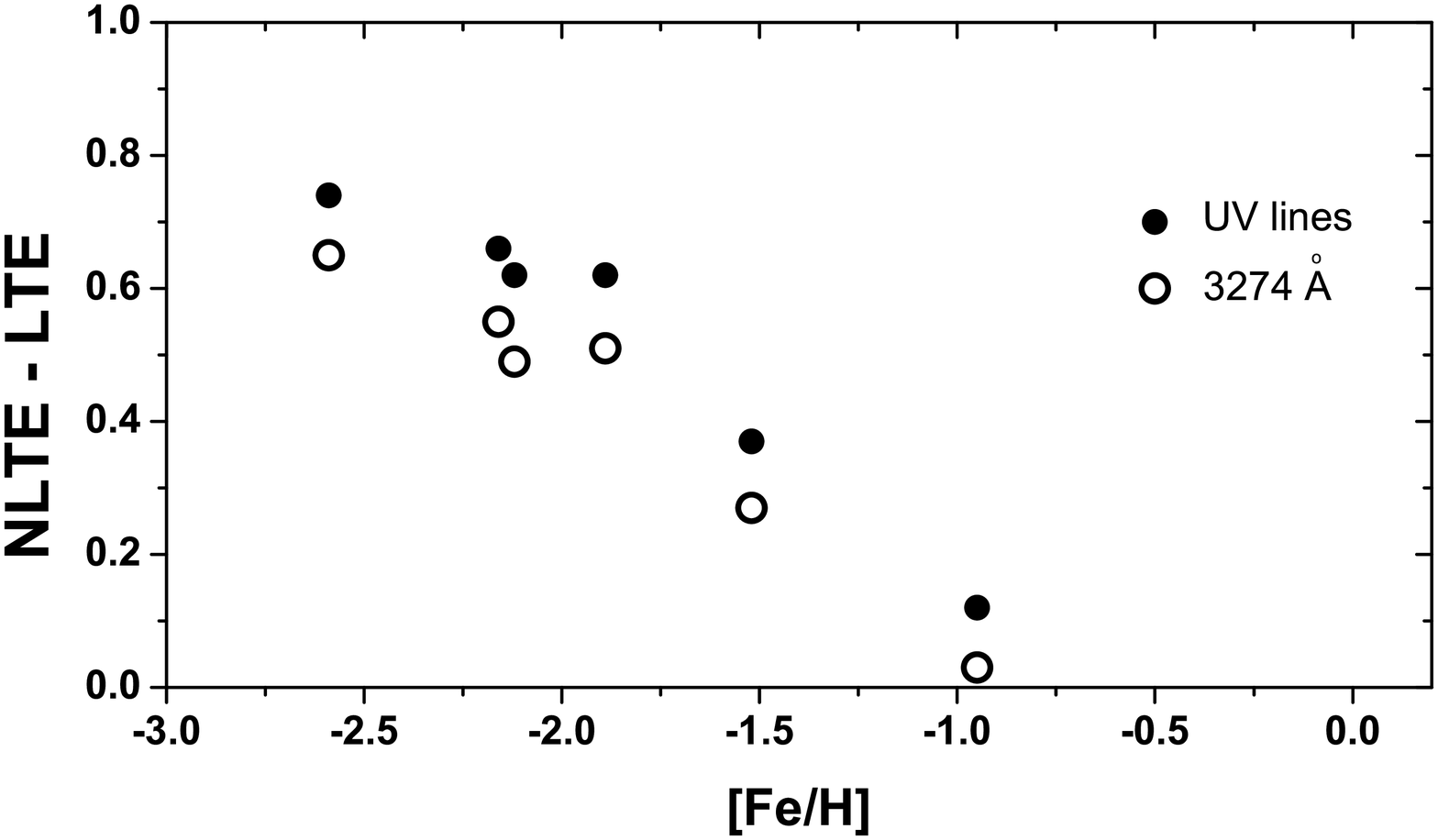}
\caption{
Non-LTE corrections to the LTE copper abundances calculated
for \ion{Cu}{i} UV lines and 3274\,\AA\ as a function of [Fe/H].
}
\label{dNLTE}
\end{figure}

What is important  is to note that all UV \ion{Cu}{i} produce very close 
NLTE-corrections. Fig. \ref{dNLTE} shows the NLTE corrections for our program 
stars (open circles -- UV lines, close circles -- resonance lines).
As we noted this recently in our paper (\citealt{Andr18}), and  as 
it also was reported by \citet{Shi14} and \citet{Yan15}, the NLTE corrections are 
close to zero for the stars of solar metallicity and quickly increase for the 
metal-poor stars.

\begin{figure}
\includegraphics[width=0.9\columnwidth,clip=true]{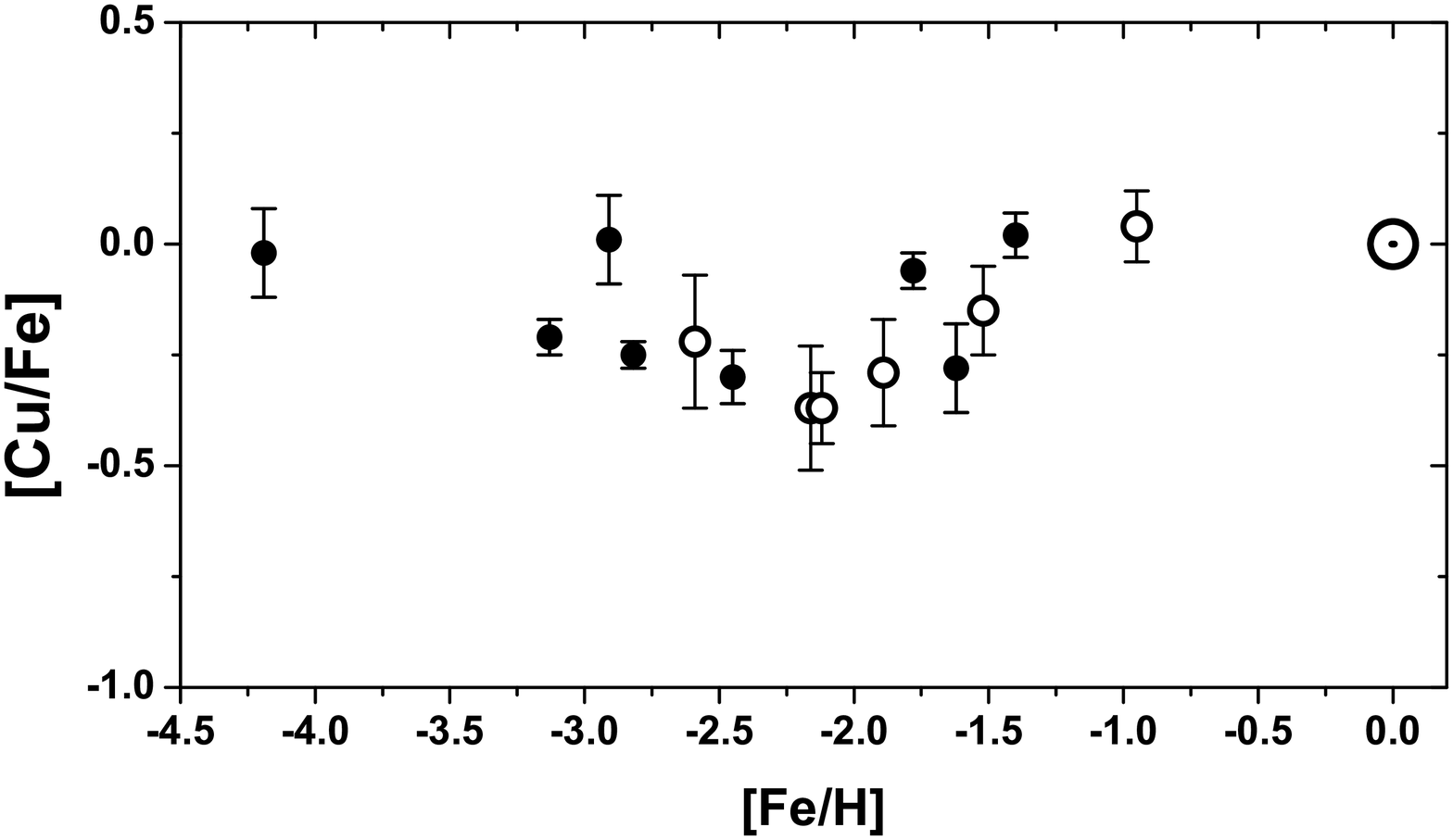}
\caption{The Galactic evolution of Cu, as captured by our programme stars.
Open symbols are our present NLTE results, filled symbols are the results from
\citet{Andr18}.}
\label{cu_fe}
\end{figure}

Fig. \ref{cu_fe} shows the copper abundance versus metallicity. Our present 
results are shown by the open circles, while those of \citet{Andr18} -- by the 
filled circles. 
Our sample consists of the main sequence stars. In the work of 
\citet{Andr18} we considered mainly cool giants. What is important to note we 
do not see any significant differences in abundances between these two samples 
(see Fig. \ref{cu_fe}). Perhaps this is the result of a small number of 
investigated objects.

We have a very good agreement between our present results based on analysis of 
NLTE \ion{Cu}{i} lines and LTE \ion{Cu}{ii} lines, and results obtained by \citet{Andr18}. 
Fig \ref{cu_fe} clearly shows that copper does not behave as a secondary 
element (as it follows from LTE data, see, e.g. \citet{Romano07} compillation 
in their Fig. 1), but on contrary it behaves as a primary element like, for instance, 
magnesium, see \citet{Andr10}, or calcium, see \citet{Spite12}. One can 
aslo note a clear difference between copper and zinc.
For example, if one supposes that copper is a $s$-process element, and behavies 
as a secondary element, then how to explain the  completely different behaviour of zinc 
(observed data cited by \citealt{TWW95}, Fig. 35, and \citealt{Romano10}, Fig. 16), 
which seems to be primary elemet as it follows from observations?  

A feature of  the copper abundance distribution (some kind of depression at about 
[Fe/H] = --2) is seen in Fig.1, and it may be caused by the beginning of era of  
the extra iron production by SNe Ia. The subsequent classic $s$-process in the 
low mass stars at higher metallicities increases again the Cu/Fe ratio. 

\section{Conclusion}

We finish our paper with the short conclusions:

1. Our NLTE consideration of the neutral copper spectrum removes disagreement 
between the copper abundance from Cu I and Cu II lines presented in
\citet{RB18}.

2. Taking into account our NLTE results on the copper abundance in stars
with metallicity ranging from --4 to --1 one can conclude that copper 
generally behaves as a primary element (at least in the metallicity
domain from --4 to --2.5, although additional data would help to
make this conclusion more convincing).

3. Since at the early stages of the Galaxy evolution ([Fe/H] from --4.5 to --3.0)
SNe II dominated as the main ISM polluters, one can suppose that the
high copper abundace comes exactly from this source of explosive nucleothynthesis.

4. If SNe II is really a main source of the copper nuclei production at the early stage 
of the Galaxy evolution, then our NLTE data on the copper abundance require significant
reconsideration of the copper yeilds in the explosive nucleothynthesis.

5. There is a hint that [Cu/Fe] ratio experiences some subsidence in its distribution
vs. metallicity at approximately [Fe/H] = --2. This could be a sign of the beginnig era 
of the ISM iron pollution from SNe I.

\section*{Acknowledgements}
SAK and AVZh acknowledge the support from the RFBR and the 
Council of Ministry of the Republic of Crimea, grant N18-42-910007.
The authors thank the anonymous referee for the offering important
suggestions.



\bibliographystyle{mnras}
\bibliography{Cu_HST} 








\bsp	
\label{lastpage}
\end{document}